# Orbital angular momentum induced beam shifts

N. Hermosa*[a], M. Merano[a], A. Aiello[b], J.P. Woerdman[a]
[a]Huygens Laboratory, Leiden University, P.O. Box 9504, 2300 RA Leiden, The Netherlands;
[b]Max Planck Institute for the Science of Light, Günther-Scharowsky-Straβe 1/Bau 34, 91058 Erlangen, Germany

## ABSTRACT

We present experiments on Orbital Angular Momentum (OAM) induced beam shifts in optical reflection. Specifically, we observe the spatial Goos-Hänchen shift in which the beam is displaced parallel to the plane of incidence and the angular Imbert-Fedorov shift which is a transverse angular deviation from the geometric optics prediction. Experimental results agree well with our theoretical predictions. Both beam shifts increase with the OAM of the beam; we have measured these for OAM indices up to 3. Moreover, the OAM couples these two shifts. Our results are significant for optical metrology since optical beams with OAM have been extensively used in both fundamental and applied research.

## 1. INTRODUCTION

It is well established that a bounded beam upon reflection and transmission on a planar interface differs in propagation with plane waves due to diffraction corrections. This may manifest as beam shifts with respect to the geometric optics prediction when reflected or refracted. The more dominant shifts are the Goos-Hänchen[1] (GH) shift in which the beam is displaced parallel to the plane of incidence, and the Imbert-Fedorov[2] (IF) shift in which the shift is perpendicular. Moreover, it has been shown that each of these two beam shifts can be separated into a spatial and an angular shift.[3] The main distinction between spatial and angular shifts is the enhancement of the latter with the propagation of the beam.

Angular Goos-Hänchen (AGH) shifts and angular Imbert-Fedorov (AIF) shifts occur only in the case of partial reflection.[3] Fig.1 shows a schematic representation of these shifts. The centroid of the reflected beam is deflected with a small angular deviation with respect to the geometric optics center. These small deviations, described by Aiello and Woerdman[3] as AGH and AIF shifts (as opposed to the conventional spatial shifts), can drastically affect the measured shifts in the position of the beam under certain experimental conditions, such as when the beam is focused. In this case, the centroid of the beam's excursion increases as the beam propagates. Merano et al.[4] showed experimental proof of the AGH effect while a quantum version of the AIF effect - the spin Hall effect of light - was demonstrated by Hosten and Kwiat[5].

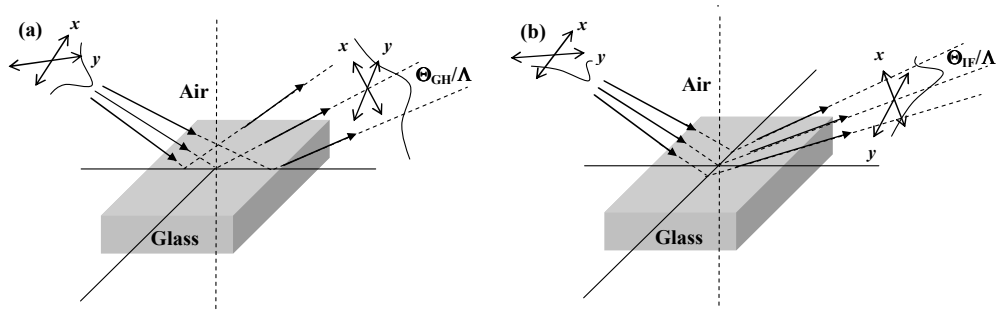

Figure 1: Schematic of the (a) Angular Goos-Hänchen and (b) Angular Imbert-Federov shifts when a beam reflects from air to glass. $\Theta_{GH}/\Lambda$ and $\Theta_{IF}/\Lambda$ are the deflection with respect to the geometric optics center (in radians), $\Lambda = kL$ where $k$ is the wave number and L is the Rayleigh length of the beam.

*hermosa@molphys.leidenuniv.nl

Recently, the GH and IF shifts were extended theoretically for the case of beams with Orbital Angular Momentum (OAM).[6,7] Fedoseyev[6] calculated spatial and angular IF shifts that depended on the beam's OAM. Bliokh[7], on the other hand, described the OAM-induced coupling between angular and spatial GH and IF shifts. Experimentally, Dasgupta and Gupta[8] verified the OAM-induced IF shift predicted by Fedoseyev but only for the spatial case.

We presented the complete four OAM-induced beam shifts- spatial GH and IF, and there angular cases, for $\ell = 1, -1$ in our recent publication.[9] In the present paper, we report for the first time experimental measurements of the OAM-induced spatial GH shift angular and IF shift for a beam with higher $\ell$. We show that these shifts are coupled by the OAM and increased proportionally to $\ell$. We compare this to our theoretical predictions derived directly from Snell's law and Fresnel equations. Also, we verify the angular nature of the IF shift by measuring the deflection of the beam as we move the detection distance.

## 1.1 Theory

The electric field of an LG beam for the case of p = 0 is given by,

$$u_{\ell 0}^{LG}(r, \varphi, z) \propto \exp\left(i\frac{k}{2}\frac{r^2}{z-iL}\right) r^\ell \exp(i\ell\varphi) \tag{1}$$

where $r, \varphi, z$ are the coordinates, $\ell$ is the azimuthal mode index, $k$ is the wave number, $L = \frac{1}{2}k\omega_0^2$ is the Rayleigh range, and $\omega_0$ is the beam waist. The reflected electric field is calculated based on the fact that the reflecting surface acts upon each plane wave in the incident field with different Fresnel coefficients. This results in a complex shift, with an imaginary part corresponding to an angular shift ("tilt"), and a real part corresponding to a spatial shift.[3] The resulting lateral and transverse displacements of the beam are combinations of both angular and spatial contributions given by;

$$\langle X \rangle(z, \ell) = \frac{1}{k}\left[\Delta_{GH}(\ell=0) - 2\ell\Theta_{IF}(\ell=0) + \frac{z}{L}(1+2|\ell|)\Theta_{GH}(\ell=0)\right] \tag{2}$$

$$\langle Y \rangle(z, \ell) = \frac{1}{k}\left[\Delta_{IF}(\ell=0) - 2\ell\Theta_{GH}(\ell=0) + \frac{z}{L}(1+2|\ell|)\Theta_{IF}(\ell=0)\right] \tag{3}$$

where $\frac{\Delta_{IF}(\ell=0)}{k}$ and $\frac{\Delta_{GH}(\ell=0)}{k}$ are the transverse and lateral spatial shifts for a Gaussian beam, respectively, and $\frac{1}{\Lambda}\Theta_{IF}(\ell=0)$ and $\frac{1}{\Lambda}\Theta_{GH}(\ell=0)$ are the transverse and lateral angular shifts also for a Gaussian beam, respectively. For a complete derivation of these quantities, we refer you to references 3 and 9. Though the calculation method was different from the one employed by Bliokh[7], the results obtained were similar, except for the factor of '2' which is due to our use of a quadrant detector.

## 2. EXPERIMENT

Laguerre-Gaussian (LG) beams were used in the experimental setup shown in Fig. 2. LG beams have well-defined OAM equivalent to $\ell\hbar$ per photon.[10] To be able to produce such beams, an open cavity HeNe laser (632.8nm) was used and forced to oscillate in higher order Hermite-Gaussian modes with a 40μm-diameter wire at the axis of the beam. These beams were then converted into higher-order LG modes by two cylindrical lenses as implemented by Beijersbergen et al[10]. Lens 1 and 2 were used for mode-matching in such a way that the beam left lens 2 collimated. Lens 3 focused the beam into a waist of $\omega_0 = 19\mu m$ to enhance its angular spread. The beam was reflected by a prism glass (BK7, n = 1.51) supported by a stage whose rotation angle can be controlled. The beam's excursion in the transverse and lateral positions was detected by a calibrated quadrant detector (New Focus 2901) when the input polarization was periodically varied between $45^0$ and $-45^0$ with a liquid crystal variable retarder. A lock-in amplifier was used to reduce technical noise.

We measure the angular nature of the shift in a separate experiment using the same setup. Here, the angle of incidence was held constant while the distance of the quadrant detector to the position of the focus was varied.

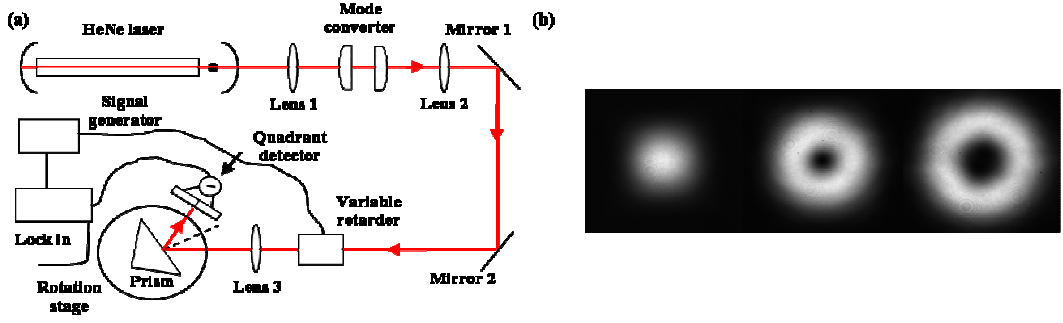

Figure 2. (a) The experimental setup. (b) LG beams produced before passing through Lens 3 (left to right, $LG_{00}$, $LG_{01}$, $LG_{02}$)

## 3. RESULTS AND DISCUSSION

We measured the beam shifts with linearly polarized beams where we switched the polarization between $45^0$ and $-45^0$; and under external reflection. Hence, eqns. 2 and 3 reduce to the following expressions;

$$\langle X \rangle(z, \ell) = \frac{-2\ell \Theta_{IF}(\ell = 0)}{k} \quad (4)$$

$$\langle Y \rangle(z, \ell) = \frac{z}{kL}(1 + 2|\ell|)\Theta_{IF}(\ell = 0) \quad (5)$$

where

$$\Theta_{IF}(\ell = 0) = \frac{\cot\theta}{R_P^2 + R_S^2}(R_P^2 - R_S^2), \quad (6)$$

$r_A(\theta) = R_A(\theta)\exp(i\phi_A)$ is the Fresnel coefficient, $A \in \{P, S\}$; and $\theta$ is the angle of incidence.

    Figure 3 shows a plot of the polarization differential transverse shifts when the polarization of the beam is switched between $45^0$ and $-45^0$. No fit parameter was used in the plots. The AIF or the angular tilt at small angles was determined by dividing the differential transverse shift with the distance between the quadrant detector and the waist of the beam. At any given angle of incidence, the ratios of the shifts with different $\ell$'s have values almost the same as the ratios of the factor $1 + 2|\ell|$, with the slight difference attributed to the difficulty of producing perfect higher order LG modes.

    Note also that in Fig. 3 that there was no difference between the positive and negative values of $\ell$ in the AIF.

    We verified the angular nature of the AIF by measuring the differential transverse shifts when the distance of the QD was increased. In Fig. 4, the linearly increasing value of the differential transverse spatial shifts showed the angular nature of the IF shifts. At a polarization state of $45^0$, the centroid of the beam was deflected with respect to the geometric optics center. It was oppositely deflected in the case of a polarization state of $-45^0$. At small angles, the difference between these deflections should increase linearly, as was observed. The geometry of this has been illustrated in Fig. 5. The two linear fit lines did not cross at zero distance, as they should do theoretically. Again, we ascribed the discrepancy to imperfections in our experimentally realized LG modes.

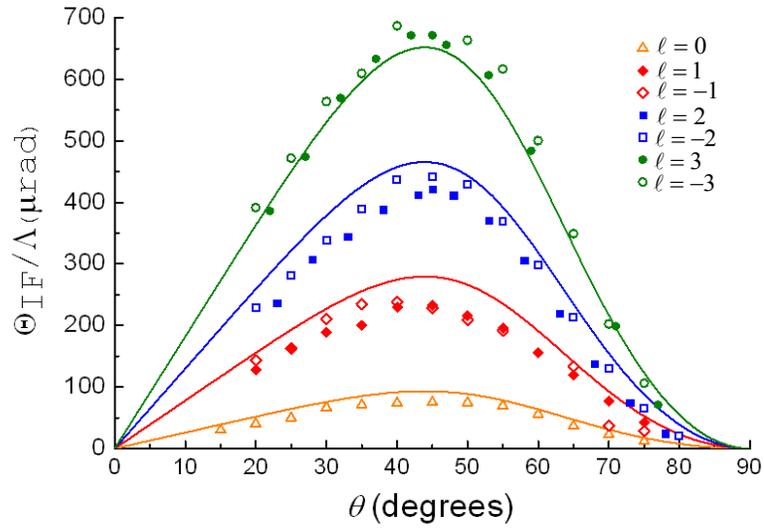

Figure 3. Angular IF shifts $\Theta_{IF}/\Lambda$ (radians, $\Lambda = kL$) of beams with different azimuthal mode index $\ell$ as a function of the angle of incidence $\theta$. The polarization was switched between $45^0$ and $-45^0$. The solid lines represent the theoretical curves. The filled and unfilled shapes of data points represent the positive and negative signs of $\ell$, respectively. We used $\Delta$ for $\ell = 0$, $\Diamond$ for $\ell = 1$, $\square$ for $\ell = 2$ and o for $\ell = 3$.

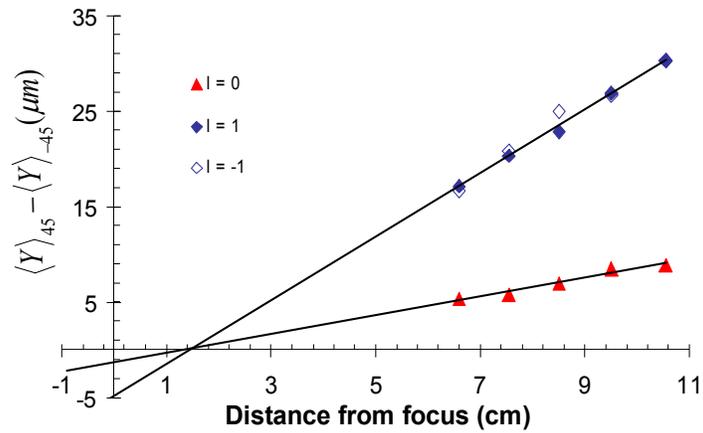

Figure 4. Transverse shifts at a constant angle of incidence $\theta$, as the distance of the detector is changed. The lines are linear fits. As the distance of the quadrant detector is moved farther from the position of the focus, the beam's transverse excursion increases linearly proving that the shift is angular in nature.

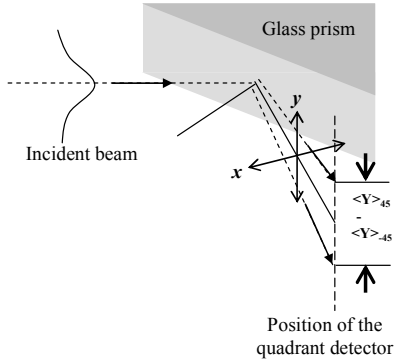

Figure 5. Transverse shifts as the distance of the detector is changed. The excursions of the beam for $45^0$ and $-45^0$ polarization are symmetric along the axis of propagation.

The spatial Goos-Hänchen shift is zero for a planar dielectric interface for a non-OAM carrying beam in external reflection geometry. In our previous publication[9], we show experimentally that this is not the case for an OAM endowed beam for the case $|\ell|=1$. The maximum shift happens when the polarization is at $45^0$.[9] Fig. 6 is a plot of the spatial GH shift from $\ell=-3$ to $\ell=3$ at a constant angle of incidence of $45^0$ which shows the linear dependence of the GH shift with $\ell$ values. The line is the theoretical fit from eqn. 4 and corresponds very well with the experimentally obtained data. A schematic diagram of the spatial GH shifts is shown in Fig. 7 as the polarization is switched.

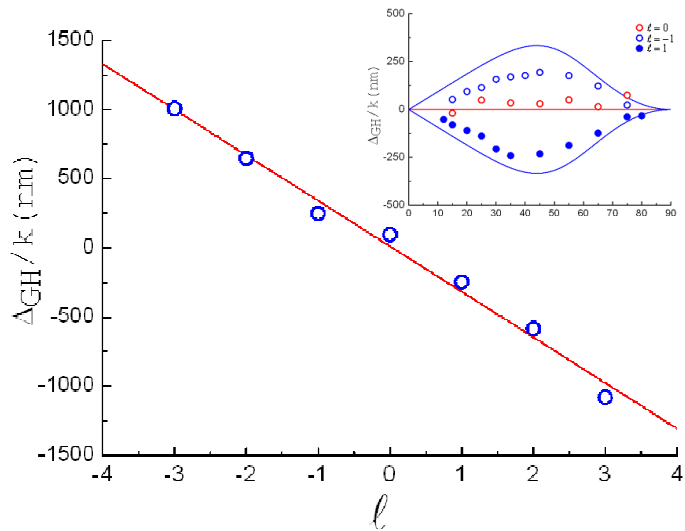

Figure 6. OAM-induced Goos-Hänchen shift at a constant angle of incidence ($45^0$). The shift is linear with increasing $\ell$ values of the beam. The inset shows the GH shift for $\ell=-1,0,1$ at different angle of incidence.

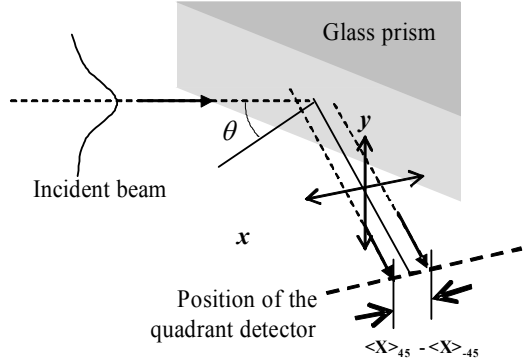

Figure 7. Schematic of the spatial GH shift. This shift on a planar dielectric in external reflection happens only for OAM carrying beams because of the coupling brought about by the OAM.

The spatial GH shift comes from the same factor responsible for the angular IF shift as seen in eqns 4 and 5. The OAM of the beam couples this factor, such that there is a spatial GH shift observed in planar dielectric surface even in external reflection. In the experiment, the AIF and the spatial GH (as shown in the inset) have the same shape except for $\ell = 0$ where the GH is zero. From experimental data on both the AIF and spatial GH shifts, we can deduce the value of $\Theta_{IF}(\ell = 0)$ by a simple multiplication of experimental parameters, $\frac{kL}{1+2|\ell|}$ and $\frac{k}{2\ell}$ for $\ell \neq 0$, for the AIF and spatial GH, respectively. The values of $\Theta_{IF}(\ell = 0)$ from AIF and spatial GH shifts are within 5% difference with each other on average which proves the same origin of the effect.

## 4. CONCLUSION

We observe the spatial Goos-Hänchen (GH)shift in which the beam is displaced parallel to the plane of incidence, and the angular Imbert-Fedorov (AIF) shift which is a transverse angular deviation from geometric optics prediction. Both beam shifts are seen to increase with the OAM of the beam. Moreover, the OAM couples these two shifts. Our experimental and our theoretical predictions agree well.

The AIF shift of a beam with an OAM has been measured for $\ell = 0, 1, 2,$ and $3$, when the beam reflects from a planar dielectric material. It is observed that the magnitudes of the shift increased with $\ell$ but are not influenced by the sign of $\ell$. The angular nature of the AIF is also verified experimentally.

Moreover, the spatial GH shift is seen to be linearly dependent on $\ell$ and reverses its sign when the sign of $\ell$ is flipped. The spatial GH shift in external reflection happens only because of the OAM of the beam.

## 5. ACKNOWLEDGEMENTS

Our work is part of the Foundation for Fundamental Research of Matter (FOM). It is also supported by the European Union within FET Open-FP7 ICT as part of the STREP Program 255914 PHORBITECH. A.A. acknowledges support from the Alexander von Humboldt Foundation.